\documentclass[12pt]{article}
\usepackage{amssymb,psfig,epsfig}
\def\gappeq{\mathrel{\rlap {\raise.5ex\hbox{$>$}} {\lower.5ex\hbox{$\sim$}}}}
\def\lappeq{\mathrel{\rlap{\raise.5ex\hbox{$<$}} {\lower.5ex\hbox{$\sim$}}}}
\def\beq{\begin{equation}} 
\def\eeq{\end{equation}} 
\def\bea{\begin{eqnarray}}
\def\eea{\end{eqnarray}}
\def\bq{\begin{quote}} 
\def\eq{\end{quote}}

\def\nn{\nonumber}
\def\ep{\epsilon}
\def\epp{\epsilon'}

\begin{document} 
\pagestyle{empty}  
\begin{flushright}  SACLAY-T01/066
\end{flushright}  
\vskip 2cm    

\begin{center}  
{\LARGE $\tau \rightarrow \mu \gamma$ and $\mu \rightarrow e \gamma$
as probes of neutrino mass models\\}
\vspace*{5mm} \vspace*{1cm}   
\end{center}  
\vspace*{5mm} \noindent  
\vskip 0.5cm  
\centerline{\bf St\'ephane Lavignac, Isabella Masina and
Carlos A. Savoy \footnote{E-mails: lavignac@spht.saclay.cea.fr,
masina@spht.saclay.cea.fr, savoy@spht.saclay.cea.fr.}}
\vskip 1cm
\centerline{\em Service de Physique Th\'eorique \footnote{Laboratoire
de la Direction des Sciences de la Mati\`ere du Commissariat \`a
l'\'Energie Atomique et Unit\'e de Recherche Associ\'ee au CNRS (URA 2306).},
CEA-Saclay}
\centerline{\em F-91191 Gif-sur-Yvette, France}
\vskip2cm 
  
\centerline{\bf Abstract}  

We discuss the possibility of discriminating between different
supersymmetric see-saw models by improving the experimental sensitivity
to charged lepton flavour violating processes. Assuming a hierarchical
neutrino mass spectrum, we classify see-saw models according
to how the hierarchy $\Delta m^2_{\odot} \ll \Delta m^2_{atm}$ is
generated, and study the predictions of each class for the branching
ratios of $\tau \rightarrow \mu \gamma$ and $\mu \rightarrow e \gamma$.
The process $\tau \rightarrow \mu \gamma$ is found to be a particularly
promising tool to probe the fundamental see-saw parameters, and
especially to identify the origin of the large atmospheric mixing angle.
Predictions for $\mu \rightarrow e \gamma$ are more model-dependent.
We point out that, even with an improvement of the experimental
sensitivities by three orders of magnitude, both $\tau \rightarrow \mu \gamma$
and $\mu \rightarrow e \gamma$ could escape detection in models where
$\Delta m^2_{atm}$ is determined by one of the lightest right-handed
neutrinos.

\vskip .3cm

\vfill
\eject 
  
  
\newpage  
\setcounter{page}{1} \pagestyle{plain}

The smallness of neutrino masses with respect to charged fermion masses
and the by now robust observation of at least one  large mixing in the
leptonic sector \cite{atmospheric,solar,SNO} suggest that the mechanism
responsible for the generation of neutrino masses is different from the
one at work for charged fermions. Presumably neutrinos are Majorana
particles and their mass is linked to lepton number violation. The
see-saw mechanism \cite{seesaw} emerges as the most elegant explanation
of the smallness of neutrino masses. At least in the supersymmetric
case, it is also very suggestive as the scale of lepton number
violation appears to be near the gauge coupling unification scale.

With three right-handed neutrinos the light neutrino effective mass
matrix then reads \footnote{We adopt the convention of writing Dirac
mass terms as $ {\bar R} m_D L$. CP violating phases will be neglected.}
\bea
{\cal M}_{\nu} & = & Y^T~ {M_R^{-1}} ~Y~ v^2_u \nn\\
& = &  U^* ~\mbox{diag}(m_1,m_2,m_3)~ U^\dagger \label{Mnu}\ ,
\eea
where $Y$ and $M_R$ are $3 \times 3$ matrices in flavour space
representing respectively the neutrino Dirac Yukawa couplings and the
Majorana right-handed mass matrix, $v_u$ is the appropriate Higgs vacuum
expectation value and $U$ is the neutrino mixing matrix.  Unless
otherwise stated we always work in the flavour space basis where the
mass matrix of charged leptons and $M_R$ are both diagonal, so that
$M_R=\mbox{diag}(M_1,M_2,M_3)$ with $|M_1|<|M_2|<|M_3|.$  

It is clear that, even if we precisely knew ${\cal M}_\nu$ by measuring $U$
and the light neutrino masses, it would not be possible to disentangle from it
the structure of the Dirac and Majorana mass matrices \footnote{ A detailed
discussion of these ambiguities is presented in \cite{Casas01}. }.  For
instance, the knowledge of $U$ does not provide direct access to the mixings
which are present in $Y$.  With $Y= V_R \mbox{diag}(Y_1,Y_2,Y_3) V_L^\dagger$,
the identification $V_L =U $ holds in a very special case, namely, if
$YY^\dagger$ and $M_R$ can be simultaneously diagonalised.  Even for small
right-mixings in $V_R$ and/or a small hierarchy among right-handed Majorana
masses, $U$ can be significantly different from $V_L$.  So, contrary to what
happens in the quark sector, neutrino masses do not provide direct access to
the left-mixings in the Yukawa couplings.  Nevertheless, as discussed later,
one can already identify phenomenologically acceptable patterns for $Y$ and
$M_R$ from the data on ${\cal M}_\nu$.

The aim of this letter is to investigate which additional information could 
be extracted from improving the  bounds on - or measuring - the branching
ratios of   the charged lepton flavour violating (LFV) processes
$\tau \rightarrow \mu \gamma$ and $\mu \rightarrow e \gamma$, in the  
framework of supersymmetric models. At present these bounds are 
$BR(\tau \rightarrow \mu \gamma) < 1.1 \times 10^{-6}$ \cite{tau_mu_gamma} 
and $BR(\mu \rightarrow e \gamma) < 1.2 \times 10^{-11}$ \cite{mu_e_gamma}.
In the Standard Model, such flavour violating processes in the charged lepton
sector are predicted to be much below the experimental bounds and negligible
with respect to the analogous ones  occurring in the quark sector, due to the
smallness of neutrino masses \cite{Petcov77}. On the other hand, one of the
major open problems of low-energy supersymmetry is to justify the fact that
flavour violating processes have not already been found at a rate
substantially larger than predicted by the Standard Model. Indeed, some amount
of flavour dependence is generically expected in the soft mass terms. Then
loops containing sleptons give rise to LFV because of lepton flavour mixing in
the  slepton mass matrices \cite{Gabbiani96,Hisano99,Feng00}. 
These non-diagonal mass matrix elements are in turn 
strongly bounded by the experimental limits on LFV transitions.

It has been pointed out \cite{Borzumati86} that if right-handed
neutrinos are present, LFV processes might get drastically enhanced with
respect to the SM --  even in the case of universality of soft SUSY
breaking terms at the scale $M_U \sim M_{Pl}$. Indeed, the
running of the slepton masses from $M_U$ to the scale where
right-handed neutrinos decouple, due to loop corrections involving the
Yukawa coupling matrix $Y,$ induces flavour non-diagonal terms.  These
terms are associated with a different combination of $Y$ and $M_R$ than
the one occurring in the see-saw, namely with the non-diagonal entries 
of the matrix
\beq
C\ =\ Y^\dagger ~\ln \left({M_U/M_R}\right)~ Y\ .
\label{ylogy}
\eeq
Thus one immediately realizes that LFV effects arising from these
loops potentially contain additional information on $Y$ and $M_R .$

In this context it has been shown recently that, also in connection
with the MSSM parameter space preferred by the present data on the muon
anomalous magnetic moment \cite{a_mu}, some particularly simple realizations
of the see-saw mechanism predict the decay of $\tau \rightarrow \mu \gamma$
and $\mu \rightarrow e \gamma$ at a rate which could be at hand of future
experimental sensitivity \cite{athand}. In particular, it turns out
that non-observation of $\mu \rightarrow e \gamma$ already excludes some of
those models in sizeable regions of the supersymmetric parameter space 
\cite{excl,Casas01}, while at present $\tau \rightarrow \mu \gamma$ is less
constraining. 

Here we take a different approach to analyse the information on the 
mechanism generating neutrino masses that could be obtained by improving the
experimental sensitivity on those branching ratios. Indeed, we adopt a
``bottom-up'' approach, i.e. we do not assume any particular model or flavour
structure for the seesaw parameters. Firstly we obtain the upper bounds on the
matrix elements of (\ref{ylogy}) from the upper bounds on $\tau \rightarrow
\mu \gamma$ and $\mu \rightarrow e \gamma ,$ respectively. Notice that the
discovery and measurement of these branching ratios would only provide upper
bounds on these one-loop corrections. Indeed, there could be other sources of
LFV in the slepton mass matrices, like flavour-dependent soft terms at tree
level mentioned above, or radiative corrections in the context of grand
unification\footnote{For instance it is well known that in $SU(5)$ an
additional source of LFV are the Yukawa couplings of the colored triplet
\cite{Barbieri95}.} and other possible contributions from unknown
physics between $M_{GUT}$ and $M_{Pl}$.  As usual we assume that these
effects do not conspire to cancel each other since they seem to have
different origins, so that each one is constrained by the experimental
bounds. 

Secondly we divide the $Y$ and $M_R$ matrices consistent with light
neutrino data into three classes which are shown to have different
patterns for (\ref{ylogy}), hence different predictions for LFV
processes.  Then it is  straightforward to see whether the planned
future searches of LFV will be sensitive enough to test such
predictions.  We find that in almost all of the mSUGRA parameter space
below the TeV region $\tau \rightarrow \mu \gamma$ appears to be a clear tool
to learn whether $Y$ possesses a small $\mu \tau$  mixing, analogous to the
quark mixing, or a large one as the light neutrinos.  This distinction is far
from being academic because, depending on it, the mechanism generating
neutrino masses has completely different characteristics which can in turn
give useful hints on the underlying flavour symmetries. In particular, the
$\tau \rightarrow \mu \gamma$ branching ratio is suppressed if the 
atmospheric neutrino oscilation mass scale is not linked to the heaviest
right-handed neutrino, as in models where the lepton analog of the CKM angles
are small. On the contrary, even if perfectly measured, the process $\mu
\rightarrow e \gamma$ alone could not achieve this task, though it would be
useful to obtain further informations on the structure of $Y$ and $M_R$ in
special situations \cite{Casas01} .

In the MSSM with right-handed neutrinos, the amplitude for the process
$l_i \rightarrow l_j \gamma$ has been completely calculated in
\cite{Hisano99}. Its dominant contribution, which scales as $\tan \beta$ for 
large $\tan \beta,$ arises from loops where charginos and sneutrinos
circulate, in analogy with the supersymmetric contribution to the muon
anomalous magnetic moment, but with an insertion of the off-diagonal element
of the sneutrino mass matrix ${m^2_{\tilde{\nu}}}_{ij}$ (always in the basis
where charged leptons and right-handed Majorana neutrinos are diagonal). The
size of this amplitude crucially depends on the mechanism of supersymmetry
breaking.  In order to pick out the effect induced by the neutrino Yukawas
$Y$, let us assume universality among soft scalar masses at $M_U \sim M_{Pl}$,
like in mSUGRA. Then the dominant contribution can be written as a function of
$m_{\tilde{\nu}}$ and $M_2$, which are respectively the (mean) sneutrino and
charged gaugino masses, $\tan \beta$ and the corresponding insertion
${m^2_{\tilde{\nu}}}_{ij}$. Below $M_1$ the off-diagonal element $(ij)$ 
of the sneutrino mass matrix is approximately given by 
\beq
{m_{\tilde{\nu}}}_{ij}^2\ \sim\ - \frac{1}{8 \pi^2} (3 m_0^2+A_0^2)  Y^*_{ki}
\mbox{ln}\left(\frac{M_U}{M_k}\right) Y_{kj}~~, \label{m2ij} 
\eeq    
$m_0$ and $A_0$ being respectively the universal soft scalar mass and
trilinear coupling at $M_U$.

\begin{figure}[!ht]
\centerline{\psfig{file=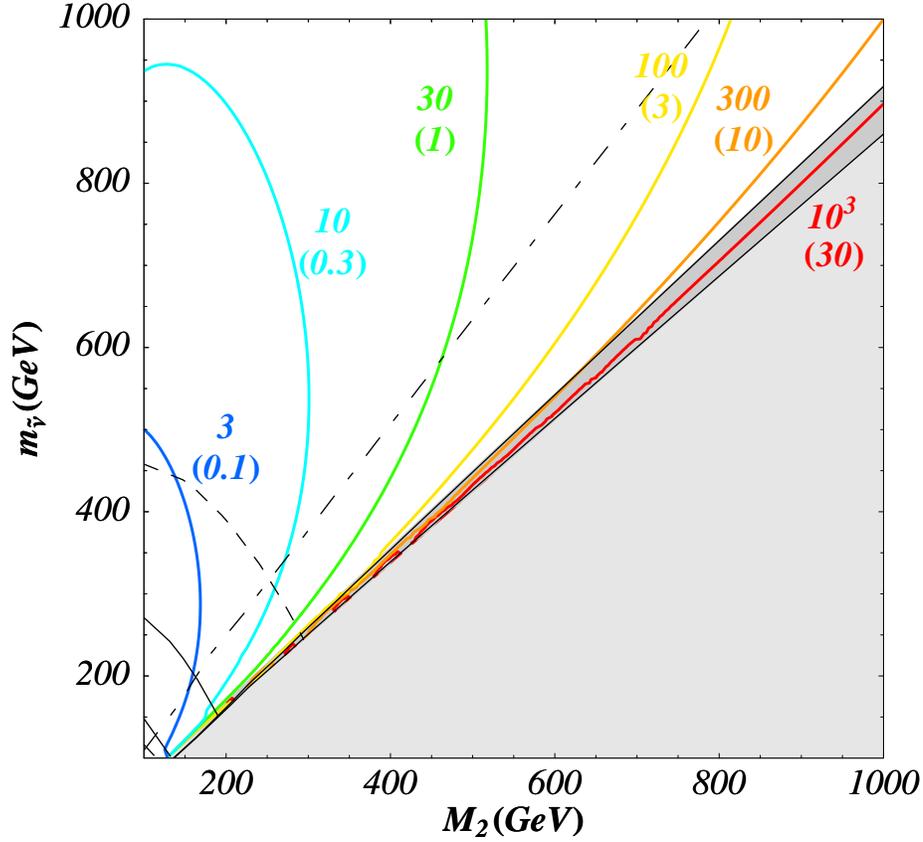,width=0.8\textwidth}}
\caption{Upper limit on the combination  $C_{\tau \mu}=
(Y^\dagger \mbox{ln}(\frac{M_U}{M_R}) Y)_{\tau \mu}$
from the present bound $BR(\tau \rightarrow \mu \gamma)<1.1 \times 10^{-6}$.
The numbers in brackets correspond to an improvement by a factor $10^3$ in
the upper bound. Here $\mbox{tan}\beta=10$, $A_0=0$ and sign$(\mu)=+$. The
solid and dashed curves in light grey represent respectively the $1\sigma$ and
$2\sigma$ contours allowed for the supersymmetric contribution to the muon
anomalous magnetic moment. The dark gray region is excluded because the
lighter stau would here be the LSP. The region below the dash-dotted line
could be explored by searching for  the LFV decay $\widetilde \chi^0_2
\rightarrow \widetilde \chi^0_1 \mu \tau$ at the LHC \cite{Hinchliffe01}.}
\label{figtmg}
\end{figure}

\begin{figure}[!ht]
\centerline{\psfig{file=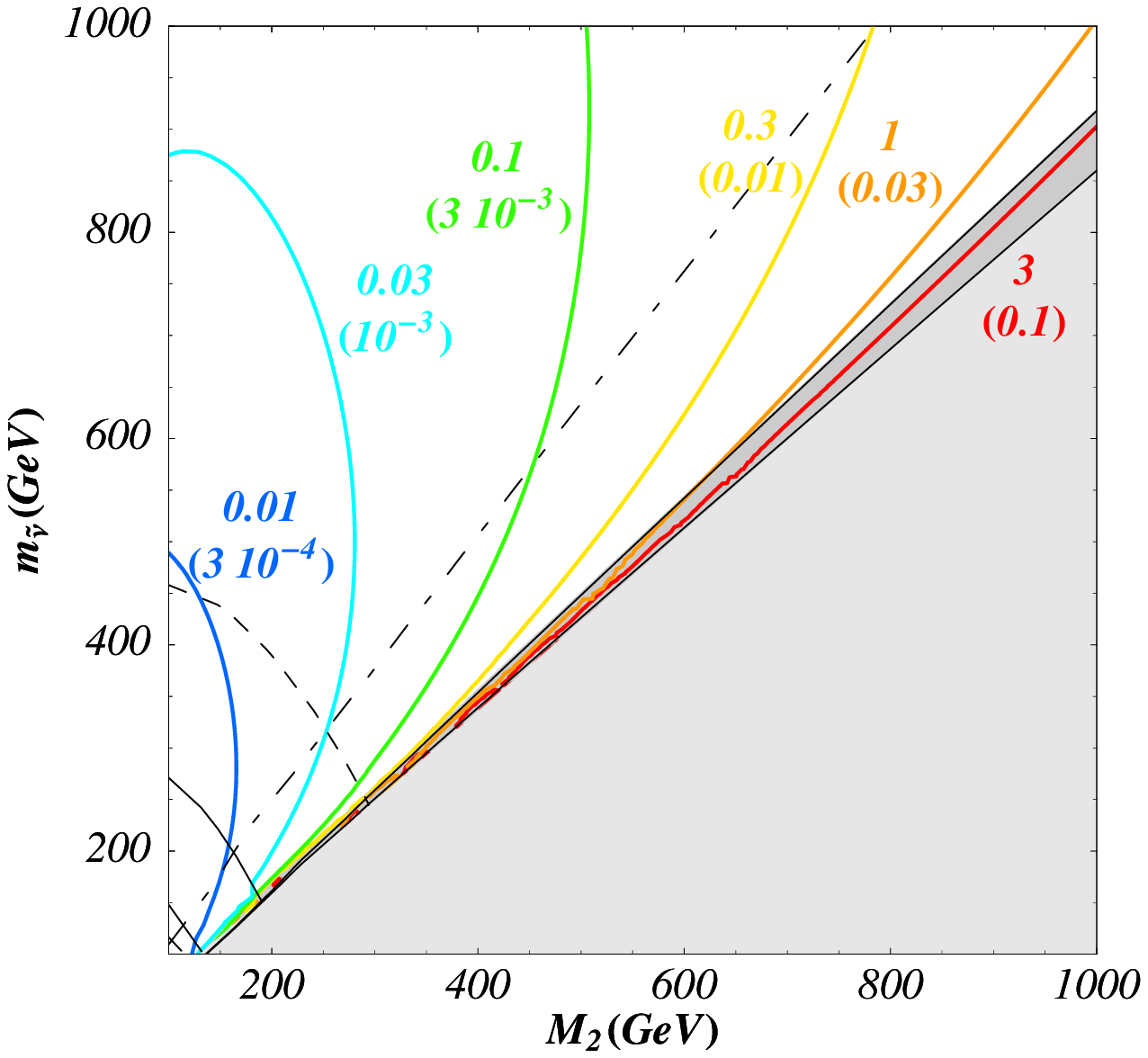,width=0.85\textwidth}}
\caption{Upper limit on the combination 
$C_{\mu e}=(Y^\dagger \mbox{ln}(\frac{M_U}{M_R}) Y)_{\mu e}$ from 
the present bound on $BR(\mu \rightarrow e \gamma)< 1.2 \times 10^{-11}$.
The numbers  in brackets correspond to an improvement by a factor $10^3$
in the upper bound. Like in fig. 1, $\mbox{tan}\beta=10$, $A_0=0$ and
sign$(\mu)=+$.}
\label{figmueg}
\end{figure}

An experimental upper limit on $\mbox{BR} (l_i \rightarrow l_j \gamma)$
gives an upper limit on the $C_{ij}$ defined in (\ref{ylogy}) for each point in
the plane $(m_{\tilde{\nu}}, M_2).$  In figs. \ref{figtmg} and \ref{figmueg} 
we show the upper limits on $C_{\tau\mu}$ and $C_{\mu e}$ as inferred from
the present bounds on the branching ratios for $\tau \rightarrow \mu \gamma$
and $\mu \rightarrow e \gamma$ respectively. We also give in brackets the
limits corresponding to an improvement by three orders of magnitude in the
sensitivity to these branching ratios. Such an improvement is indeed
expected for $\mu \rightarrow e \gamma$ \cite{futdirexp}. Prospects for $\tau
\rightarrow \mu \gamma$ are currently less optimistic, but searches for
the LFV decay $\widetilde \chi^0_2 \rightarrow \widetilde \chi^0_1 \mu \tau$
at future colliders could provide limits on $C_{\tau\mu}$ of that order of
magnitude in the region of the $(m_{\tilde{\nu}}, M_2)$ plane indicated on
figs. \ref{figtmg} and \ref{figmueg} \cite{Hinchliffe01}. 

The results in the figures are displayed in the plane
$(m_{\tilde{\nu}}, M_2)$, the variables which they are mostly sensitive to.
They have been determined in the framework of the universality assumption of
the mSUGRA model, where the MSSM parameter $\mu$ basically depends on the
gaugino mass universality, while the insertions ${m_{\tilde{\nu}}}_{ij}^2$ as
given by (\ref{m2ij}) depend more on the universality assumption within the
slepton and Higgs sector.  It is then easy to estimate the change in the
results for reasonable departures from these assumptions.

In these figures we put $\tan \beta =10$ but, since for large
$\mbox{tan} \beta$ the bounds on the $C_{ij}$ scale as the inverse of
$\mbox{tan}\beta$, the upper bounds for other $\mbox{tan} \beta$ values
are obtained upon multiplication by $(10/\tan\beta)$.

As stressed before, since light neutrinos and LFV processes provide us
with complementary information on the fundamental see-saw parameters,
it is interesting to understand which level of improvement in the
experimental sensitivity to LFV processes would make it possible to
discriminate between different classes of models and to shed some light
on the underlying flavour theory.  It is therefore a useful task to
classify realistic see-saw models of neutrino masses according to their
predictions for the branching ratios of LFV processes - or more
exactly, given our ignorance of the supersymmetric mass spectrum, for
the quantities $C_{ij}$.

Neutrino oscillation experiments provide data on $U$, $\Delta m^2_{atm}
= m^2_3-m^2_2$ and $\Delta m^2_{\odot} = m^2_2-m^2_1.$ In the absence
of a direct mass measurement, three possibilities are still open for
the light neutrino spectrum:  (A) $|m_3| \gg |m_{2}| \geq |m_{1}|$, (B)
$|m_1|\simeq |m_2| \gg |m_3|$ and (C) $|m_1|\simeq |m_2| \simeq |m_3|$. A
hierarchical pattern with a dominant third family is the case for up quarks,
down quarks and charged leptons.  The almost perfect degeneracy of squared
masses required by (B) and (C) could only result from a conspiracy between the
Dirac and the Majorana mass matrices.  Moreover, a precise degeneracy at the
unification scale would be stabilized against radiative corrections down to
low energies only if it were protected by a symmetry \cite{RGE_nu}.

In the following we thus focus on case (A), namely a hierarchical
spectrum for light neutrinos.  We make the following assumptions about
the neutrino mass spectrum: the atmospheric neutrino anomaly is mainly
due to $\nu_\mu \leftrightarrow \nu_\tau$ oscillations with a large
mixing angle, as suggested by the Super-Kamiokande \cite{atmospheric} and
CHOOZ \cite{CHOOZ} data; the solar neutrino deficit \cite{solar,SNO} is
accounted for by one of the allowed solutions: MSW-SMA, MSW-LMA, LOW
\footnote{ At the time when this letter was completed, no analysis of
solar neutrino oscillations including the SNO result had been published
yet. The VO solution ($\Delta m^2 \lesssim \mbox{a few}\, 10^{-10}\, eV^2$)
was disfavoured by 3-flavour global analyses of all neutrino data
\cite{Gonzalez-Garcia01}. The 2-flavour analyses of solar neutrino data
that appeared just after the SNO result also disfavour the SMA solution
\cite{fit_including_SNO}.}. There is no oscillation explanation of the (still
controversial \cite{KARMEN}) LSND result \cite{LSND}. Note that due to the
hierarchy of masses, the only remaining constraint from terrestrial
experiments is the CHOOZ limit on $|U_{e3}|^2$. Then $m_3^2 \simeq \Delta
m^2_{atm} = (1.4-6.1) \times 10^{-3} \mbox{eV}^2 ,$ $|U_{\mu 3}| / |U_{\tau
3}| \approx 1 ,$ and $|U_{e3}| < .2 .$ \cite{Gonzalez-Garcia01}. Concerning 
$\Delta m^2_{\odot}$ and the solar angle, given by $\tan
\theta_{12}=|U_{e2}|/|U_{e1}|$, one has to keep in mind all possible solar
solutions.

This strongly constrains the effective neutrino mass matrix ${\cal
M}_\nu$.  In particular its entries must be such that the two heaviest
eigenstates are strongly mixed while being well separated in mass.
Without any further assumption, ${\cal M}_\nu$ can be written in terms
of the parameter $\ep \equiv m_2 / m_3$:
\beq
{\cal M}_\nu\ =\ m_3 \left( U_{\tau 3}^2\left( \begin{array}{ccc} 
b^2 & ab & b \\ ab & a^2 & a \\ b & a & 1  \end{array} \right)\ + 
{\cal O} (\ep ) \right)~~ ,
\label{structure}
\eeq
where $a \equiv U_{\mu 3} / U_{\tau 3}$, $ b \equiv U_{e 3}/U_{\tau 3}$
and the symbol ${\cal O} (\ep )$ stands for a matrix whose elements are
at most of order $\ep$ and whose structure is linked to the solar
neutrino angle.  As discussed above, $|U_{\tau 3}|=0.71\pm 0.15\,
,\ |a| \approx 1\, ,\  |b| <.28\, ,$ and the central values for $\ep$
are $0.1,\ 0.04$ and $0.006$ for the LMA, SMA and LOW solutions, 
respectively \cite{Gonzalez-Garcia01}, for $m^2_1 \ll m^2_2$.

The expected branching ratios for LFV processes crucially depend on how
the structure (\ref{structure}) is realized in terms of $Y$ and $M_R$.
In order to be able to classify the various possibilities, let us have
a closer look at the lower right $2 \times 2$ submatrix in ${\cal
M}_\nu$, henceforth denoted by ${\cal M}^{(\mu ,\tau )}_\nu .$ Clearly
the structure (\ref{structure}) corresponds to the relations:
\beq
x \equiv \frac{\det {\cal M}^{(\mu ,\tau )}_\nu}
{\left(\mbox{Tr}{\cal M}^{(\mu ,\tau )}_\nu \right)^2} \lesssim {\cal O}
(\ep )~~~,
~~~~~~({\cal M}_\nu)_{\mu \tau} \approx ({\cal M}_\nu)_{\tau \tau}~~~.
\label{det/tr}~~~
\eeq
This is automatically satisfied if one of the three right-handed neutrinos,
say $N_k$, gives the dominant contribution to each entry in ${\cal
M}^{(\mu ,\tau )}_\nu$ \cite{King_B439,AF_JHEP},
\beq
{\cal M}^{(\mu ,\tau )}_\nu\ =\ v^2_u \sum_{k=1}^{3}
\left( \begin{array}{cc}\frac{Y^2_{k\mu}}{M_k}
& \frac{Y_{k\mu}Y_{k\tau}}{M_k} \\
\frac{Y_{k\mu}Y_{k\tau}}{M_k}
& \frac{Y^2_{k\tau}}{M_k} \end{array} \right)~~,  
\label{dominance}
\eeq
and if $Y_{k\mu} \approx Y_{k\tau}$. If this is not the case, some amount
of cancellation among the entries of $Y$ is needed in order to fulfil
the first condition in (\ref{det/tr}).

We shall refer to the situation where the contributions of the other
eigenstates of $M_R$ are at most ${\cal O} (\ep )$ with respect to
those of $N_k$ as ``dominance of $N_k$''.  According to this definition,
we divide the see-saw models leading to the structure (\ref{structure})
into three different classes that we turn to discuss in connection with
their implications for radiative LFV decays. A few typical examples for
each class, chosen among the many models in the literature, are aimed to
illustrate its general features.

\noindent {\bf [Class 1:]} {\it None of the three right-handed neutrinos
dominates in} ${\cal M}^{(\mu , \tau)}_\nu$. 

This case arises when all $\frac{Y^2_{k\mu}}{M_k}$ and
$\frac{Y^2_{k\tau}}{M_k}$ are of the same order of magnitude or, for at
most one value of $k$, are smaller by at least a factor $\ep$.
Since $Y_{k\mu} \sim Y_{k\tau}$, the atmospheric mixing is
automatically large.  This, together with the hierarchy condition in
(\ref{det/tr}), requires a tuning of these $Y$ couplings. Depending on
the solar neutrino solution this tuning may be mild (LMA) or more
severe (LOW).  Generically one expects a large $\mu\tau$ mixing in
$Y$, $(V_L)_{\mu\tau} = {\cal O}(1)$ -- in contrast to the analogous
mixing in the quark sector -- unless there is a strong hierarchy
in $M_R$, $M_3 \gg M_2/\ep $ .

This class of models leads to \footnote{This is only a lower limit
because if the heaviest right-handed neutrino $N_3$ is not dominant in
${\cal M}^{(\mu , \tau)}_\nu$, its contribution to $|C_{\tau \mu}|$ may
increase the result in (\ref{class1}) by a factor at most ${\cal O}
(\ep M_3 /M_2 )$.}
\bea
|C_{\tau \mu}|\ \gtrsim\ |Y_{k\tau}|^2\, \ln \left( \frac{M_U}{M_k} \right)\ 
\sim\ \frac{\sqrt{\Delta m^2_{atm}}}{v_u^2}M_k\ln \left( \frac{M_U}{M_k} 
\right) \sim \left( \frac{M_k}{5 \times 10^{14}\mbox{GeV}}\right) 
\ln \left( \frac{M_U}{M_k} \right)~~ ,
\label{class1}
\eea   
where $M_k$ is the largest $M_R$ eigenvalue corresponding to a dominant
contribution.  For $ Y_{k\tau} \sim {\cal O} (1) ,$ so that $M_k \sim 5
\times 10^{14} \mbox{GeV} ,$ $|C_{\tau \mu}| \gtrsim 7 .$ We can
therefore conclude that, for  $Y_{k\tau}^2 \tan \beta \geq 30$ and
$M_2,m_{\tilde \nu} \leq 1$ TeV, $\tau \rightarrow \mu \gamma$ should be
observed provided that the experimental sensitivity improves by three
orders of magnitude (see fig. \ref{figtmg}).  For $Y_{k\tau}^2 \tan
\beta < 30$ there remains a narrow band at large values of
$M_2,m_{\tilde \nu}$, corresponding to the cosmologically preferred
region of mSUGRA \cite{Ellis01}, where $\tau \rightarrow \mu \gamma$
could escape detection.

The predictions for $\mu \rightarrow e \gamma$  and $\tau \rightarrow
e\gamma$ are more model-dependent. If the heaviest eigenstate $N_3$
gives one of the dominant contributions in ${\cal M}_\nu^{(\mu,\tau)}$,
then  $|C_{\tau e}| \sim |C_{\mu e}| \gtrsim |Y_{3e}C_{\tau
\mu}/Y_{3\tau}| .$ In this case an upper bound of $10^{-14}$ in $\mu
\rightarrow e \gamma$ would imply $Y_{3e} \lesssim 10^{-2} $ (see fig.
\ref{figmueg}).

Models based on Abelian flavour symmetries \cite{FN} fall
into class 1, unless the charges are chosen in such a way
that texture zeros due to the holomorphy of the  superpotential, 
appear in suitable entries of $Y$, as in examples 2 and 3 below.

{\it Example 1.} Consider an Abelian horizontal (or flavour)
symmetry $U(1)_X$ broken by a single MSSM singlet field $\theta$
\footnote{The conclusions are identical with any number of Abelian
symmetries and an equal number of ``flavons''.}.  The charges of the
leptons are denoted by $l_i$, $e_i$ and $n_i$ for the doublet leptons
$L_i$, right-handed charged leptons $E_i$ and right-handed neutrinos
$N_i$ respectively.  The charges of the Higgses are denoted by $h_u$
and $h_d$. If thee are no supersymmetry zeros, the $n_i$ factorize out in
the see-saw mechanism  implying that all right-handed neutrinos contribute to
${\cal M}^{(\mu,\tau)}_\nu$ so that these models belong to class 1. The large
atmospheric mixing angle requires $l_2=l_3$, and one ends up with a light
neutrino mass matrix of the form \cite{Irges98} : 
\beq  
 {\cal M}_\nu\ \sim\
 \frac{v^2_u\, \lambda^{2(l_3+h_u)}}{M}   \left( \begin{array}{ccc}
   \lambda^{2(l_1-l_3)} & \lambda^{l_1-l_3} & \lambda^{l_1-l_3} \\ 
   \lambda^{l_1-l_3} & d & e \\
   \lambda^{l_1-l_3} & e & f
   \end{array} \right)~~ ,
\label{eq:M_nu_example_1}
\eeq
where the scale $M$ is related to the breaking of lepton number,
$\lambda=\theta /M ,$and the coefficients $d$, $e$, $f$ in ${\cal M}^{(\mu
,\tau )}$ are $O(1).$  The condition (\ref{det/tr}) requires a tuning of 
$x = \frac{df-e^2}{(d+f)^2}$ to ${\cal O}(\ep )$. If $x \lesssim
\lambda^{l_1-l_3} $, one finds $m_2 \sim \lambda^{l_1-l_3}\, m_3$ and $\Delta
m^2_{\odot} / \Delta m^2_{atm} \sim \mbox{max}(x  \lambda^{l_1-l_3},
\lambda^{3(l_1-l_3)})$, and a mixing matrix $U$ with all elements of ${\cal
O}(1)$ with the exception of $U_{e3} \sim \lambda^{l_1-l_3} .$ We thus obtain
a large solar mixing angle.  In principle this allows us to reproduce either
the large angle MSW solution for typical values $x \sim \lambda^{l_1-l_3}
\sim 0.1$  (with $|U_{e3}|$ close to its present upper bound) or the
LOW solution. However the latter option requires a strong tuning, $x
\sim \lambda^{l_1-l_3} \sim 5 \times 10^{-3}$.  At the price of an even
stronger tuning larger values of $U_{e3} \sim \lambda^{l_1-l_3}$ are in
principle possible.

We have in this kind of example, $|C_{\tau \mu}|\sim\ |Y_{3\tau}|^2
\ln\left(M_U/M_3\right)$ and $|C_{\mu e}|/|C_{\tau \mu}|\sim\lambda^{l_1-l_3}
.$ Thus, for $Y_{3 \tau} \sim 1$, $|C_{\tau \mu}|\sim 7$ and  $|C_{\mu e}|
\sim 0.7, 0.04$ for LMA and LOW respectively.  We conclude that $\mu
\rightarrow e \gamma$ will allow us to test or exclude this model if the
solution of the solar neutrino problem is LMA.  For LOW the analysis is less
conclusive if $Y^2_{3 \tau} \tan \beta < 30$.  These conclusions agree with
those obtained in previous studies \cite{athand,Casas01}.  However they are
related to the absence of texture zeros in $Y$ and/or $M_R$ and are not general
amongst the class 1 models.

\noindent {\bf [Class 2:]} {\it The heaviest right-handed neutrino $N_3$
dominates in} ${\cal M}^{(\mu , \tau)}_\nu$.

This is realised when the hierarchy in $Y Y^T$ for the $(\mu , \tau)$
sector is stronger than the one in $M_R$ by at least ${\cal O}(\ep^{-1}) .$
A mass hierarchy is naturally obtained
in this case and $Y_{3\mu} \approx Y_{3\tau}$ is needed in order to have
a large mixing; hence the large mixing must be already present in $Y$, that
is $(V_L)_{\mu \tau} = {\cal O}(1)$.  In this case the coefficient of the
$\tau \rightarrow \mu \gamma$ amplitude is
\beq 
 |C_{\tau \mu}|\ \approx\ |Y_{3\tau}|^2\, \ln \left( \frac{M_U}{M_3} \right) \,
 \approx  \left( \frac{M_3}{5 \times 10^{14}\mbox{GeV}}\right) 
 \ln \left( \frac{M_U}{M_3}\right)\ ,
\eeq
so that this class of models always predicts a sizeable branching ratio
for $\tau \rightarrow \mu \gamma$. In particular
$ |C_{\tau \mu}| \sim 7 $ if $Y_{3\tau} \sim {\cal O}(1)$.
The branching ratio for $\mu \rightarrow e \gamma $ is more
model-dependent as it depends on the hierarchy between the $Y_{ke}$
couplings. More precisely there is a lower bound $|C_{\mu e}/C_{\tau
\mu}| \gtrsim |Y_{3e}/Y_{3\tau}| ,$ while $|Y_{3e}/Y_{3\tau}| \lesssim
\mbox{max}(b,\ep) .$

In order to ensure the stronger hierarchy in $Y$, a richer flavour
structure than for class 1 has to be at work
\cite{AF_B451,King99,AFM_SU5,Berezhiani01}.

{\it Example 2.} As an example of dominance of $N_3$, let us consider
an Abelian flavour symmetry with the following charge assigment
\footnote{This charge assignement corresponds to a slight change in the
model of Ref. \cite{AFM_SU5}, which has a small value of $\tan \beta$.}:
$(l_1,l_2,l_3)=(2,0,0)$, $(n_1,n_2,n_3)=(1,-1,0)$, $(e_1,e_2,e_3)=(3,2,0)$,
$h_d=1$ and $h_u=0.$ The small symmetry breaking parameter is taken to be the
Cabibbo angle, $\lambda \simeq 0.22$.  The characteristic of the model with
respect to the previous example is the existence of supersymmetric zeros in
$Y$ and $M_R$ in the basis of the Abelian charge eigenvalues. One then
obtains, in the basis of charged lepton and right-handed neutrino mass
eigenstates: 
\beq
  Y \sim \left( \begin{array}{ccc}
    \lambda & \lambda & \lambda  \\
    \lambda & \lambda & \lambda  \\
    \lambda^2 & 1 & 1  \end{array} \right)~~ ,  \quad
  M_R \sim M \mbox{diag}(-1, 1, 1)~ , \quad
  {\cal M}_\nu \sim \frac{v^2_u}{M} \left( \begin{array}{ccc}
    \lambda^4 & \lambda^2 & \lambda^2  \\
    \lambda^2 & 1 & 1  \\
    \lambda^2 & 1 & 1
  \end{array} \right)~~ .
\label{ex2}
\eeq
The $-1$ in $\mbox{diag}\, (-1, 1, 1)$ indicates that $N_1$ and $N_2$ are
quasi degenerate in mass, with opposite $CP$ parities.  As discussed above,
the large mixing in the atmospheric neutrino sector originates from $Y_{32}
\approx Y_{33}$, and the right mass scale is obtained for $M \sim 5 \times
10^{14}\,$ GeV.  One can check that $N_3$ dominates in the $\mu \tau$ sector,
naturally suppressing the ratio $m_2 / m_3$.  Here $\Delta m^2_{\odot}$ is
further suppressed relative to $\Delta m^2_{atm}$ by the fact that $m_1$ and
$m_2$ are almost degenerate, giving rise to the LOW solution.  As for LFV
processes, from the structure of $Y$ in (\ref{ex2}) one immediately realises
that the contribution of $N_3$ is dominant in $\tau \rightarrow \mu \gamma$,
while all three right-handed neutrinos give a contribution of similar size to
$\mu \rightarrow e \gamma$: $ |C_{\tau\mu}|\sim 7$ and $|C_{\mu e}|\sim
\lambda^2 \ln (M_U/M_3)\sim 0.4.$ We conclude that an improvement of the
experimental limit on $\mu \rightarrow e \gamma $ by three orders of magnitude
will test this model for $M_2,m_{\tilde\nu} \leq 1\, \mbox{TeV}.$ Notice that
the same model with $h_d=0$, corresponding to $\tan \beta \sim 50$, is almost
excluded by the present bounds on $\mu \rightarrow e \gamma.$

\noindent {\bf [Class 3:]} {\it One of the two lightest right-handed neutrinos
dominates in} ${\cal M}^{(\mu , \tau)}_\nu$. 

This is realised when the hierarchy in $Y Y^T$ for the $(\mu , \tau)$
sector is weaker than the one in $M_R$ by at least ${\cal O}(\ep^{-1}) .$
Like in class 2 models, a mass hierarchy is
naturally obtained, and the condition for a large mixing is $Y_{k\mu}
\approx Y_{k\tau}$, where $N_k$ is the dominant right-handed neutrino.
Unlike class 2 models and the major part of class 1 models
this is not incompatible with a hierarchical structure of the Dirac
mass matrix.  Hence the large $\mu \tau $ mixing does not require a
large ${V_L}_{\mu \tau}$.  Since $m_3 \sim v^2_u\, Y^2_{k\tau}/M_k$
corresponds to the atmospheric neutrino scale, the dominance of $N_k$
implies $M_3 \gtrsim \ep^{-1} M_k Y^2_{3 \tau}/Y^2_{k \tau} \, \sim  (5
\times 10^{14}\, \mbox{GeV}) Y^2_{3 \tau}/\ep$, suggesting a high scale
for the breaking of lepton number.  Indeed, in explicit models $M_3$ is
generally larger than the GUT scale.

Smaller branching ratios for LFV processes can be expected for two reasons: 
first, since $Y$ does not necessarily contain a large $\mu \tau$ mixing, 
$Y_{3 \mu}$ can be smaller than $Y_{3 \tau}$; second, since $N_3$ is not the
dominant right-handed neutrino, $Y_{3 \mu}$ and $Y_{3 \tau}$ can be both
significantly smaller than 1 without requiring a small $M_3$. Then $N_3$  does
not necessarily provide the dominant contribution to $C_{\tau \mu}$, unlike
class 2 and the major part of class 1. 

In order to ensure the weaker hierarchy in $Y$, a richer flavour structure
than for class 1 has to be at work \cite{Shafi99,Barbieri99,AFM00}.

{\it Example 3.} As a first example, consider another variant of the
$SU(5)$ model of Ref. \cite{AFM_SU5}, namely an Abelian horizontal symmetry
with charges $(l_1,l_2,l_3)=(2,0,0)$, $(n_1,n_2,n_3)=(3,1,0)$,
$(e_1,e_2,e_3)=(4,3,1)$, $h_d=1$ and $h_u=-2$.  These charges imply a
different texture of supersymmetric zeros in $Y$ and $M_R$ with
respect to the previous example.  One then obtains, in the basis of
charged lepton and right-handed neutrino mass eigenstates:
\beq
  Y \sim \left( \begin{array}{ccc}
    \lambda^3 & \lambda & \lambda  \\
    \lambda & \lambda^3 & \lambda^3  \\
    1 & \lambda^2 & \lambda^2  \end{array} \right)~~ ,  \quad
  M_R \sim M \mbox{diag}(\lambda^6, \lambda^2, 1)~ , \quad
  {\cal M}_\nu \sim \frac{v^2_u}{M \lambda^4} \left( \begin{array}{ccc}
    \lambda^4 & \lambda^2 & \lambda^2  \\
    \lambda^2 & 1 & 1  \\
    \lambda^2 & 1 & 1
  \end{array} \right)~~ ,
\eeq
where again $\lambda \simeq 0.22$.  As can be easily checked, $N_1$
dominates in all entries of ${\cal M}_\nu$ but $({\cal M}_\nu)_{11}$. The
large mixing in the $\mu \tau$ sector is due to $Y_{12} \approx Y_{13} ,$ and
$\Delta m^2_{atm}$ yields $M \equiv M_3 \sim 2 \times 10^{17}\, \mbox{GeV}$,
well above the GUT scale. The LOW solution is naturally obtained.  As for LFV
processes, one immediately sees that $N_1$ gives the largest contribution to
$\tau \rightarrow \mu \gamma$, while the most important contribution to $\mu
\rightarrow e \gamma$ is provided by $N_3$:  $|C_{\tau\mu}|\sim \lambda^2 \ln
(M_U/M_1) \sim 0.5$ and $|C_{\mu e}|\sim \lambda^2 \ln (M_U/M_3) \sim 0.1.$
Thus in this model $\mbox{BR} (\mu \rightarrow e \gamma)$ lies within the
expected experimental sensitivity for $M_2, m_{\tilde \nu} \leq 1$ TeV, while
even with an improvement of the experimental sensitivity by three orders of
magnitude, $\tau \rightarrow \mu \gamma$ should remain unobserved in a large
region of the $(M_2, m_{\tilde{\nu}})$ plane.  For $\tan \beta = 10$, this
region lies to the right of the curve labelled by the numbers $10\, (0.3)$ in
fig. \ref{figtmg}.

{\it Example 4.} As a second example, let us consider a model
\cite{Barbieri99} based on a $U(2)$ flavour symmetry. In order to explain
the hierarchy of fermion masses, a sequential breaking of the flavour
group, involving several flavon fields, is required:  $U(2)
\stackrel{\ep}{\longrightarrow} U(1)
\stackrel{\epsilon'}{\longrightarrow} 1$, with $\epsilon' \ll
\epsilon$.  After supplementing the model with a vertical $SO(10)$
structure, a good account of the known quark and charged lepton
properties is obtained for $\ep \simeq 2 \times 10^{-2}$ and $\epsilon'
\simeq 4 \times 10^{-3}$ \cite{U2}.  In the basis where charged leptons
and heavy Majorana masses are diagonal, one obtains for this model:
\beq
  Y\ \sim\ \left( \begin{array}{ccc}
    s_E \epsilon' & \epsilon' & \epsilon' \\
    \epsilon' & \epsilon & 1 \\
    \epsilon' & \epsilon & 1
  \end{array} \right) ,  \quad
  M_R\ \sim\ M \epsilon\, \mbox{diag}\, (\epsilon \epsilon'^2, 1, 1)\ , \quad
  {\cal M}_\nu\ \sim\ \frac{v^2_u}{M \epsilon^2} \left( \begin{array}{ccc}
    s^2_E & s_E & s_E  \\
    s_E & 1 & 1  \\
    s_E & 1 & 1
  \end{array} \right) ,
\eeq
where $s_E = \sqrt{m_e/m_\mu} \simeq 0.07$. In this example, $M_1$
dominates in all entries of ${\cal M}_\nu$. The large atmospheric angle
is due to $Y_{12} \approx Y_{13}$ and the fit of $\Delta m^2_{atm}$
yields $M_3 \sim 3 \times 10^{16}\, \mbox{GeV}$ (and $M \sim 10^{18}\,
\mbox{GeV}$). The presently disfavoured SMA solution is achieved.

For LFV processes, both $M_2$ and $M_3$ contribute to $C_{\tau\mu}$ and
$C_{\mu e}$ at leading order, yielding $C_{\tau\mu}\sim \ep \ln (M_U/M_3)
\sim 0.1$ and $C_{\mu e}\sim \ep \epp \ln ( M_U/M_3)\sim 3\times 10^{-4} .$
Thus in this model, even with an improvement of the experimental sensitivity
by three orders of magnitude, both $\mu \rightarrow e \gamma$ and $\tau
\rightarrow \mu \gamma$ should remain unobserved in a large region of the
$(M_2, m_{\tilde{\nu}})$ plane.  For $\tan \beta = 10$, this region lies to
the right of the curves labelled by the numbers $3\, (0.1)$ and $0.01\, 
(3\times 10^{-4})$ in figs. \ref{figtmg} and \ref{figmueg}.

\begin{table} [p]
\begin{center}
\begin{tabular}{|c||c|c|c|}
\hline 
& & & \\ 
Pattern for ${\cal M}_\nu$ & Class 1 & Class 2 & Class 3 \\ 
& & & \\ 
\hline \hline
& & &\\
Dominant $N_k$ in ${\cal M}_{\nu}^{(\mu,\tau)}$ & none & $N_3$ (heaviest)&
$N_1$ or $N_2$ \\
& & &\\ 
\hline
& & & \\
Hierarchy of $Y Y^T$ w.r.t. $M_R$& commensurate & stronger & milder \\
& & & \\
\hline
& & \multicolumn{2}{|c|}{}  \\
Origin of  & tuning: conspiracy  
& \multicolumn{2}{|c|}{natural: hierarchy in $Y$}  \\
$\ep \equiv m_2/m_3 \ll 1$ & between $Y$ and $M_R$ & 
\multicolumn{2}{|c|}{ leads to dominance 
of one $N_k$} \\
& & \multicolumn{2}{|c|}{}  \\
\hline
& \multicolumn{2}{|c|}{} &  \\
${(V_L)}_{\mu\tau}$ ($\nu_L$ flavour mixing) &
\multicolumn{2}{|c|}{large}
&  possibly small\\
& \multicolumn{2}{|c|}{} &  \\
\hline
& & &   \\
characteristic &  Abelian without
& {Abelian with } & {non-Abelian/Abelian }  \\
flavour symmetries &texture zeros &texture zeros & with texture zeros \\
& &  & \\  
\hline
 &\multicolumn{2}{|c|}{} & \\
heaviest  & 
\multicolumn{2}{|c|}{$M_3 \sim (5 \times 10^{14}\, \mbox{GeV})\,
Y_{3\tau}^2$}  &  $M_3 > (10^{16}\, \mbox{GeV})\, Y_{3\tau}^2$ \\
Majorana mass & \multicolumn{2}{|c|}{} & $M_1 \ll M_3$  \\
&  \multicolumn{2}{|c|}{} & \\
\hline
& \multicolumn{3}{|c|}{} \\
consequences if &  \multicolumn{3}{|l|}{\hskip .2cm -- model-dependent,
related to the solar neutrino solution} \\
$BR(\mu \rightarrow e \gamma) < 10^{-14}$&
\multicolumn{3}{|l|}{\hskip .2cm -- class 3 models favoured} \\
& \multicolumn{3}{|c|}{} \\
\hline
& \multicolumn{2}{|c|}{} & \\
consequences if  &\multicolumn{2}{|c|}{most of TeV region excluded }   
& favoured for small \\
$BR(\tau \rightarrow \mu \gamma) < 10^{-9}$ &
\multicolumn{2}{|c|}{(a narrow band close to the cosmo-}      
&  $Y_{3 \mu}$ and/or $Y_{3 \tau}$ \\
& \multicolumn{2}{|c|}{logically preferred region remains)}  & \\
&  \multicolumn{2}{|c|}{} & \\
\hline
\end{tabular}
\end{center}
\vskip .5cm
\centerline{Table 1}
\label{tab}
\end{table}

Let us now compare our analysis with previous studies of LFV decays. Most of
them (see e.g. Ref. \cite{athand, excl}) adopt a ``top-down'' approach, i.e.
they consider specific models in which the structure of $Y$ and $M_R$ is more
or less fixed. As a matter of fact, these generally fall into our class 1.
Recently Casas and Ibarra \cite{Casas01} have undertaken a detailed analysis
of $\mu \rightarrow e \gamma$  based on a ``bottom-up'' approach with the
neutrino oscillation experiments as input. In practice however, they were led
to make simplifying assumptions in order to reduce the number of arbitrary
parameters in the $C_{ij}$, thus missing some interesting physical
possibilities such as the dominance of one of the two lightest right-handed
neutrinos, like in our class 3. By contrast our analysis, which uses the
``bottom-up'' approach as well, aims at classifying models of hierarchical
neutrino masses according to the typical size of their predictions for LFV
decays, stressing the relative model-independence, within a given class, of
the forcast for $BR(\tau \rightarrow \mu \gamma)$.

Finally we should mention Ref. \cite{Davidson01}, which discusses the
different problems one would confront in trying to reconstruct the
fundamental seesaw parameters $Y$ and $M_R$ from weak-scale precision
measurements including neutrino masses and mixings, the rates of LFV 
processes and an accurate determination of the sparticle mass spectrum.

In this letter, we have addressed the issue whether, depending on the
pattern of the fundamental see-saw parameters, lepton flavour violating
decays may lie within the reach of the prospective experiments or not.
Classifying the see-saw models according to the mechanism which generates the
physical parameter $\Delta m^2_{\odot} / \Delta m^2_{atm}$ and large mixing
for the atmospheric neutrino, allows to characterize each class through
physically meaningful properties such as the size of the flavour mixing in the
Dirac mass matrix and the scale of right-handed neutrino masses. Most
importantly, the proposed experiments on LFV decays could discriminate among
these classes if supersymmetric particles exist below the TeV region. This
analysis is summarized in Table 1. It is worth stressing that models where
smaller LFV effects are predicted from the mechanism discussed in this paper,
cannot be excluded from the discovery of these decays since there are many
alternative sources of flavour mixing in the supersymmetric theories. 

As a general trend, if the $\tau \rightarrow \mu \gamma $ branching ratio
turns out to be below $10^{-9}$  -- or equivalent constraints are obtained
through LFV in sparticle decays --  models where the heaviest right-handed
neutrino only marginally contributes to $\Delta m^2_{atm}$ are favoured. This
class of models allows for small mixings in the left-handed neutrino sector,
so that leptons would behave quite the same as quarks, as expected in a
grand-unified scenario. The heaviest right-handed neutrino also tends to be
close or above the GUT scale in this class of models. On the contrary,
predictions for $\mu \rightarrow e \gamma $ are very model-dependent.


\vskip 1cm
\noindent
{\bf Acknowledgements:}
this work has been supported in part by the RTN European Program
HPRN-CT-2000-00148.


\end{document}